\documentclass[aps,prl,twocolumn,superscriptaddress,showpacs,amsmath,amssymb]{revtex4-1}
\usepackage{graphicx,bbm}
\usepackage{hyperref}
\hypersetup{colorlinks=true,citecolor=[rgb]{0.09,0.36,0.55},linkcolor=[rgb]{0.2,0.63,0.17},urlcolor=[rgb]{0.09,0.36,0.55}}%
\usepackage{mathtools}
\usepackage{color}
\newcommand{\be}{\begin{equation}}
\newcommand{\ee}{\end{equation}}
\newcommand{\bea}{\begin{eqnarray}}
\newcommand{\eea}{\end{eqnarray}}

\newcommand{\un}[1]{{\underline{#1}}}

\newcommand{\ave}[1]{{\langle #1\rangle}}
\newcommand{\ii}{ {\rm i} }
\newcommand{\dd}{ {\rm d} }

\newcommand{\ZZ}{\mathbb{Z}}
\newcommand{\RR}{\mathbb{R}}

\newcommand{\oo}[1]{{[ #1 ]}}

\def\be{\begin{equation}}
\def\ee{\end{equation}}

\def\ave#1{\langle #1 \rangle}
\def\ii{{\rm i}}

\def\tit#1{}
\newcommand{\half}{{\textstyle\frac{1}{2}}}

\usepackage{color}

\begin{document}
\title{Diffusion in deterministic interacting lattice systems}
\author{Marko Medenjak}
\author{Katja Klobas}
\author{Toma\v z Prosen}
\affiliation{Faculty of Mathematics and Physics, University of Ljubljana, Jadranska 19, SI-1000 Ljubljana, Slovenia}
\begin{abstract}
  We study reversible deterministic dynamics of classical charged particles on a~lattice with
  hard-core interaction. It is rigorously  shown  that the system exhibits three types of transport
  phenomena, ranging from ballistic, through diffusive to insulating. By
  obtaining an~exact expressions for the current time-autocorrelation
  function we are able to calculate the~linear response transport coefficients,
  such as the diffusion constant and the Drude weight. Additionally, we calculate
  the long-time charge profile after an~inhomogeneous quench and obtain diffusive
  profile with the Green-Kubo diffusion constant. 
  Exact analytical results are corroborated by Monte-Carlo
  simulations.
\end{abstract}

\maketitle

{\bf Introduction.--} Understanding out-of-equilibrium phenomena has been
at the~forefront of condensed matter physics of the last decade. Despite the~efforts, we have only
recently gained a considerable insight into the~microscopic origins of
the~transport in interacting systems in terms of an emerging field of generalized
hydrodynamics (GHD)~\cite{spohn2012large,Bernard_Doyon,PhysRevX.6.041065,bertini,spyros,de2016non,ilievski2017microscopic,bulchandani2017bethe,doyon2017dynamics,doyon2017geometric,doyon2017soliton}.
While GHD provides a general~framework to analytically deal with the ballistic transport, in particular in integrable systems, it lacks an~extension which would enable us to study normal, diffusive transport. 
In this regard, there are only a~few results on lower bounding the~diffusion constant in terms of local, or bilocal conserved charges~\cite{ProsenPRE14,medenjak2017lower}.

Since the~integrable systems are characterized by ballistically propagating
excitations \cite{bonnes2014light}, the~question of how the~diffusion, which is usually
related to microscopic chaos, arises in integrable, locally interacting, clean and deterministic (i.e. non-disordered and non-dissipative) systems is puzzling to say the~least 
\cite{ljubotina2017spin,PZ09,Steinigewegdif,Robin,RobinTypicality,Fabian}. A~great deal of attention has been devoted to the~study of
inhomogeneous quench problems, where two chains in equilibrium with distinct temperatures or
chemical potentials are joined
together and then let to evolve under a homogeneous Hamiltonian~\cite{spohn2012large,Bernard_Doyon,PhysRevX.6.041065,bertini,spyros,de2016non,ilievski2017microscopic,bulchandani2017bethe,doyon2017dynamics,Spohn1977}.
In such situations, non-equilibrium steady states typically emerge on ballistic lines~$x=v t$. However, for systems exhibiting parity-like symmetries, with respect to which the initial quench state is symmetric and
the current anti-symmetric, the ballistic transport channel may close and time-dependent DMRG simulations clearly indicate that the steady state arises along diffusive lines~$ x=\xi \sqrt{t} $~\cite{ljubotina2017spin}. 

In this Letter we expound one possible general mechanism for diffusive behavior in interacting systems, which as we show
on an~example, is connected to the~interplay between freely propagating
neutral degrees of freedom and insulating behavior of charge carrying ones.
The~model in question is a~simple, reversible cellular automaton, consisting of three types
of particles: freely moving vacancies, and hard-core interacting
positive and~negative charges. Despite the~integrability, obtaining the~full time dependence is usually
intractable, except for non-interacting systems. 
Typically this presents an insurmountable barrier for calculation
of transport coefficients. In our model, however, we explicitly compute the~time dependence of
current time autocorrelation functions in separable equilibrium states, and 
solve the inhomogeneous quench problem with arbitrary initial charge density bias resulting in a universal diffusive error function scaling profile.
Depending on the~density of vacancies and the~imbalance of
positive and negative charge three different regimes are identified.
The~absence of vacancies renders the~system insulating, while in a generic case
of the charge imbalance the~system exhibits ideal transport. The~regime with a
finite density of vacancies and without the charge imbalance is especially
interesting, since it includes the~maximum entropy state, and in this regime
the~model exhibits purely diffusive transport.

\begin{figure}
  \centering
  \includegraphics[width=\columnwidth]{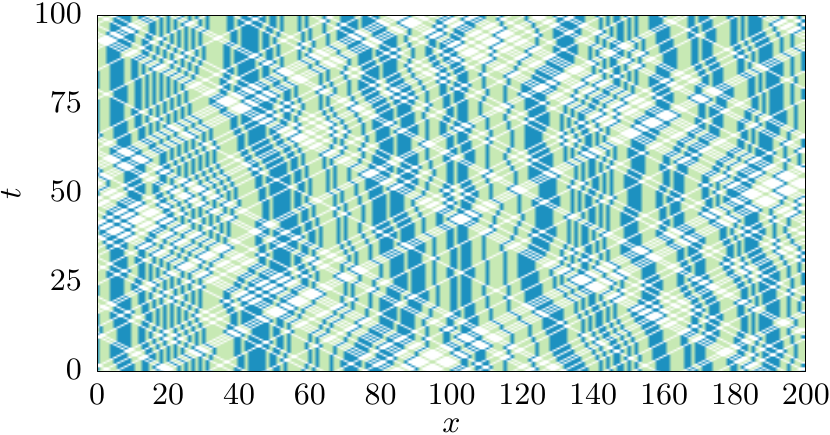}
  \caption{\label{oneconfiguration} Time evolution of a~random maximum entropy configuration ($\rho=2/3,\mu=0$) for $200$ sites. Particles~$+$, $-$, and vacancies $\emptyset$ 
  are in blue, green, and white, respectively.}
\end{figure}

{\bf The~model.--}
Consider a~deterministic, reversible cellular automaton defined on the~chain
with~ an even length of $ n $ sites. Each site can be either vacant (state $\emptyset $), or occupied by a positively or negatively charged particle (state $+$ or $-$). 
The dynamics of the lattice configuration, $\un{s}=(s_1,\ldots,s_n),s_x\in\{\emptyset,+,-\}$, can be expressed in terms of a local two site mapping $\phi_{x,x+1}(\un{s})=(s_1,s_2,\dots,s'_x,s'_{x+1},\dots,s_n)$, where the updated elements $(s_x',s_{x+1}')$ are obtained from the initial ones $ (s_{x},s_{x+1}) $, by a self-inverse transformation
\begin{equation}\label{eq:localpropagator}
  (\emptyset,\emptyset) \leftrightarrow  (\emptyset,\emptyset),\;
  (\emptyset,\alpha)  \leftrightarrow  (\alpha,\emptyset), \;
    (\alpha,\beta)  \leftrightarrow  (\alpha,\beta),
\end{equation}
with $\alpha,\beta \in \{+,-\}$. The local process describes the elastic scattering of  charged particles. The lattice configuration at time $ t ,$ $ \un{s}^t=\phi(\un{s}^{t-1}) $, can be expressed in terms of a two step propagator $\phi = \phi^{\text{o}}\circ \phi^{\text{e}}$ given by sequences of disjoint local mappings \eqref{eq:localpropagator}
\begin{equation}
 \phi^\text{o}=\phi_{1,2}\circ \dots \circ \phi_{n-1,n},\quad \phi^\text{e}=\phi_{2,3}\circ \dots \circ \phi_{n,1}.
\end{equation}
The dynamics is induced by freely propagating vacancies, while the clustered particles remain completely frozen in time as illustrated in~Fig.~\ref{oneconfiguration}.
The~cellular automaton admits a mechanical analogy in terms of a two-species, synchronous hard-point gas if particles at occupied sites $x$ are attributed velocities $2(-1)^x$ or $-2(-1)^x$, at odd or even steps, 
corresponding respectively to integer or half-integer times
\footnote{
In this sense our model belongs to a class of hard-rod systems, where diffusion has been extensively studied  \cite{jepsen1965,lebowitz1968,lebowitz1969,durr1985,VanDenBroeck}.
However, considering particle positions to take values on the lattice $\ZZ$ rather than on the line $\RR$ has two conceptual advantages: 
(i) The information (entropy) density in the initial condition is qualitatively smaller and diffusion cannot be attributed to random phases with which different particles collide even if velocities have a discrete distribution.
(ii) Dynamics can be analyzed in terms of simple linear algebra and matrix spectral problems.}.

\begin{figure}
  \centering
  \includegraphics[width=\columnwidth]{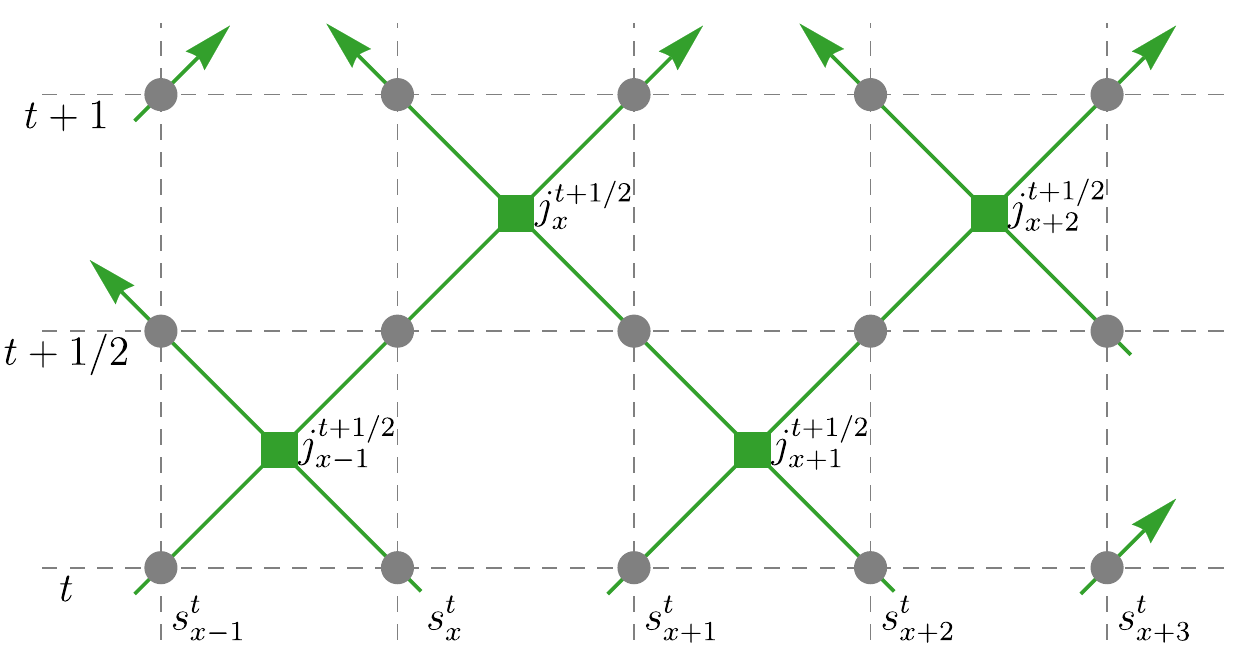}
  \caption{\label{mod}Schematic representation of the model. Gray vertices denote
  the~sites to which the particles or vacanices ($s_x^t$) are assigned, and green boxes where the particles scatter carry local
  current. Particles can be imagined to move along green lines. Pairs of sites $(x-1,x)$, $(x+1,x+2)$,\ldots are updated
  between the~time-slices~$t$ and $t+\frac{1}{2}$, while the~shifted pairs are propagated
in the~following half-time step.}
\end{figure}

We are interested in the~dynamics of the~charge~$q^t_x$ corresponding to sites~$ x $ and~$ x+1 $ at time $ t $,
\begin{equation}
  q^t_x=s_x^t+s_{x+1}^t,
  \label{gostota}
\end{equation}
with $ s^t_x=\pm 1$, if the~site $ x $ is occupied by the~positive/negative
charge, and $ s_x^t=0 $ otherwise. The~total charge $Q^t=\sum_x q^t_x$, is
a~constant of motion. The~corresponding current $ J^t=\sum_x j_x^t $
can be defined on the~intermediate time steps (see~Fig.~\ref{mod}) as 
\begin{equation}
  j_x^{t+1/2}=2\,(-1)^x (s^{t+1/2}_{x}-s^{t+1/2}_{x+1})(s^{t+1/2}_x+s^{t+1/2}_{x+1})^2,
  \label{tok}
\end{equation}
where $ j_x $ denotes the local density. The~only configurations that contribute to the~current are $ (\pm,\emptyset) $ and $ (\emptyset,\pm) $.
One can check that the~continuity equation
\begin{equation}\label{eq:continuityeq}
  q_x^{t+1}-q_x^t+\frac{1}{2}\left(j_{x+1}^{t+1/2}-j_{x-1}^{t+1/2}\right)=0
\end{equation}
is satisfied.

To carry out the~calculations we introduce a commutative algebra of observables  ${\cal A}$ (multiplicative algebra of functions over the configuration space) with a {\em local} basis 
\begin{eqnarray}
  &[\alpha]_x(\un{s})=\delta_{\alpha_x,s_x}, \quad \alpha \in  \{\emptyset,+,-\},\\
  &[\alpha_1 \alpha_2 \ldots \alpha_r]_x = [\alpha_1]_{x} [\alpha_2]_{x+1} \cdots [\alpha_r]_{x+r-1}.
\end{eqnarray}
The subscript $x$ will be dropped when the position is clear from the context. 
The dynamics of observables $a^t(\un{s})\equiv U^t a(\un{s}) = a(\phi^{t}(\un{s}))$ is given by a $3^n \times 3^n$ matrix $ U $ \footnote{note that in the canonical basis \unexpanded{$[\alpha_1,\dots,\alpha_n]$}, $ U $ is a permutation matrix given in terms of local propagators \unexpanded{$(U_{x,y})_{\un{s}',\un{s}} = \delta_{\un{s}',\phi_{x,y}(\un{s})}$}} factorized as
\begin{equation}
U = U^{\rm o} U^{\rm e},\; 
U^{\rm o} = \prod_{x=1}^{n/2} U_{2x-1,2x},\;
U^{\rm e} = \prod_{x=1}^{n/2} U_{2x,2x+1}.
\end{equation}
Note that the local propagator obeys Yang-Baxter equation $U_{x,y}U_{y,z}U_{x,y}=U_{y,z}U_{x,y}U_{y,z}$. 
Additionally, it proves useful to
introduce the~unnormalized maximum entropy state  
\begin{equation}
  \langle a\rangle=\sum_{\underline{s}} a(\underline{s}),
\end{equation}
in terms of which an~expectation value in any probability
distribution (state) $ p $ can be expressed as $\langle a\rangle_p= \langle a\, p\rangle$. 
We introduce an alternative local basis and its dual
\begin{align}
  \label{ort_b}
  [0] & =  [\emptyset]+[+]+[-], & [0]' & = (1-\rho) [\emptyset]+\frac{\rho}{2} \left([+]+[-]\right),\nonumber\\
  [1] & =  [+]-[-], & [1]' & =  \half\left([+]-[-]\right),  \\
  [2] & = \tfrac{1}{1-\rho}[\emptyset]-[0], & [2]' & = \tfrac{1-\rho}{2}\left(2[\emptyset]-[+]-[-] \right), \nonumber
\end{align}
such that $ [0]\equiv 1 $ and $\ave{[\alpha][\beta]'} = \delta_{\alpha,\beta}$, $\alpha,\beta\in\{ 0,1,2\}$. The parameter $ \rho $ will be connected to the density of particles later on.
The~local charge and current now read
\begin{gather}
  \label{eq:localchargedensity}
  q_x=
  \oo{10}_x+\oo{01}_x,\\
  j_x=  2(1-\rho) (-1)^x\,
  \left(\oo{10}_x-\oo{01}_x +\oo{12}_x
  -\oo{21}_x \right).
  \label{eq:localcurrentdensity}
\end{gather}

{\bf Linear response.--}
According to Einstein's relation the~diffusion coefficient $ \mathcal{D} $ is
connected to conductivity as~${\sigma}=\chi \mathcal{D}$, $\chi$ being
the static susceptibility (i.e.~the second moment of charge~$Q$). In the~small
constant electric field, the conductivity takes the~following form (Sect.~A of \cite{sup})
\begin{equation}\label{prev}
  \sigma=\frac{1}{2} C(0)+\sum_{t=1}^{\infty}  C(t),
\end{equation}
where $ C(t)= \lim_{n\to\infty}\frac{1}{n}\langle J U^t J\rangle_p $ is the current correlation function, with $ p $ being an~equilibrium state, $U p = p$.
Another important transport coefficient, corresponding to the~rate at which
the~conductivity diverges~\cite{IP13}, is the~Drude weight,
$D=C(\infty)$.

We restrict the~discussion to translationally invariant
product equilibrium states~$p=p(\rho,\mu)=\prod_{x=1}^n p_x$,
\begin{equation}
\label{state}
  p_x=[0]'_x+\mu [1]'_x,\qquad
  0\le \rho\le 1,\quad-\rho \le \mu \le \rho.
\end{equation}
Note that $ [0]'_x $ depends on $ \rho $.
The~density $\rho $ represents the~probability of a~lattice site being
occupied by a~charged particle, while the~chemical potential~$ \mu $
corresponds to the~charge imbalance. The static susceptibility in such a state is $ \chi=4(\rho-\mu^2) $.

\begin{figure}
  \centering
  \includegraphics[width=\columnwidth]{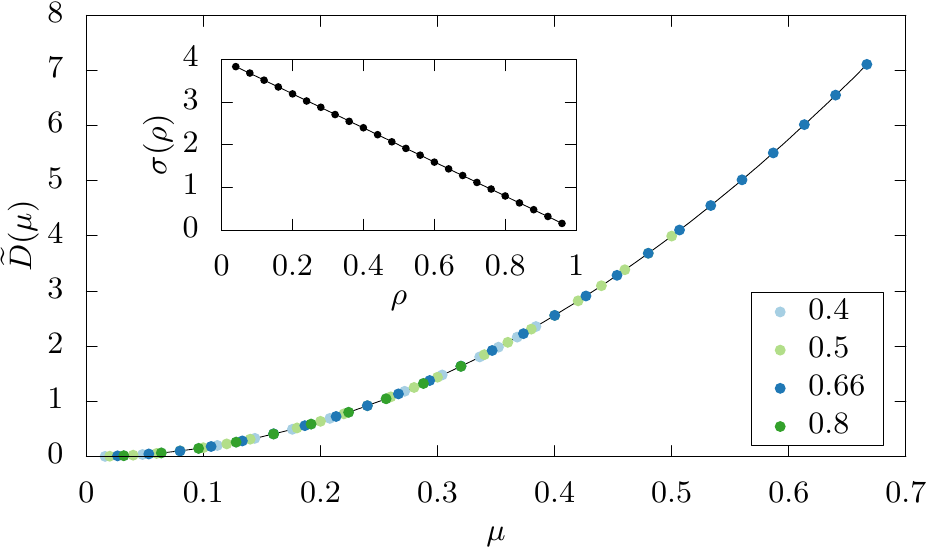}
  \caption{\label{fig:autocorr}
  The~comparison between the~exact and numerically estimated
  Drude weight~$\widetilde{D}$ and conductivity $\sigma$.
  Drude weight is rescaled as $\widetilde{D}=D\,/(\rho^{-1}-1)$.
  Points with different colors correspond to numerically calculated
  Drude weights at different values of~$\rho$ and black points (inset)
  are numerically estimated values of the~conductivity
  (at~$\mu=0$). The~black lines correspond to exact
  values~\eqref{eq:realDiffusionConstant} and~\eqref{eq:drudeWeight}.
}
\end{figure}

{\em Diffusive regime.--}
Initially we consider a balanced equilibrium state with $\mu=0$ and arbitrary $\rho$ for which the following orthogonality relations hold,
\begin{equation}
  \ave{\oo{0}\oo{1}}_p=
  \ave{\oo{0}\oo{2}}_p=
  \ave{\oo{1}\oo{2}}_p=0.
\end{equation}
As a consequence only the~observables with a~single occurrence of~$\oo{1}$
and at most one~$\oo{2}$ in the~propagated current $ J^t $  can
contribute to the~expectation value~\eqref{prev}.
The local observables that are relevant for the calculation of the diffusion constant are $y_0=\sum_x \left(\oo{10}_{2x}-\oo{01}_{2x}\right)$,
$y_1=\sum_x \left(\oo{12}_{2x}-\oo{21}_{2x}\right)$ and
$y_2=\sum_x \left(\oo{012}_{2x}-\oo{021}_{2x}\right)$, in terms of which the current reads~$J=2(1-\rho)(-2 y_0-y_1+y_2)$.
In addition, we note that two steps of the propagator are conjugated $ U^\text{o}=S^{-1}U^\text{e}S $ by a lattice shift, defined as $ S [\un{\alpha}]_x=[\un{\alpha}]_{x+1} $.
Since $ y_{\alpha} $ are translationally invariant, i.e. $ S^2 y_{\alpha}= y_{\alpha}$, the complete propagator on the relevant subspace reads $ U=(S U^\text{e})^2 $.
Under the half-time propagation~$S U^{\text{e}}$ the observables $y_0$ and $y_1$
map into linear combinations of $y_0$ and $y_2$. Due to the ballistic propagation of $ \oo{2} $, $U_{1,2}\oo{02}=\oo{20}$, $U_{1,2}\oo{20}=\oo{02}$, any additional basis operators appearing in $ y^t_2$, $t>0$, are orthogonal to $J$, since there is always at least one occurrence of $\oo{0}$ between  $\oo{1}$ and $\oo{2}$. Therefore, in order to compute $ C(t) $, we only have to consider the time evolution restricted to the subspace spanned by $\{y_\alpha\}  $.
Half-time step propagator $ S U^{\text{e}} $ projected to this basis reads  
\begin{equation}\label{eq:reducedpropagationDif}
  \mathcal{U}=\begin{bmatrix}
    1-2\rho & 2\rho & 0 \\
    0 & 0 & 0 \\
    -2(1-\rho) & 1-2\rho & 0 \\
  \end{bmatrix},
\end{equation}
and yields the~following simple expression for the~current-autocorrelation function
\begin{equation}
  \label{eq:diffusiveautocorrelation}
  \begin{aligned}
    C(t)&=4\rho(1-\rho)
    \begin{bmatrix}2(1-\rho),& \rho ,& -\rho\end{bmatrix}
    \mathcal{U}^{2t}
    \begin{bmatrix}
      2 \\1 \\-1 \\
    \end{bmatrix}\\
    &= \begin{cases}
      8\rho(2-\rho)(1-\rho);& t=0,\\
      16\rho(1-\rho)^4(1-2\rho)^{2t-2};& t\ge 1,
    \end{cases}
  \end{aligned}
\end{equation}
The row vector contains properly normalized overlaps $\langle y_\alpha J\rangle_p$.
One can then easily calculate the~conductivity and the~diffusion constant
\begin{equation}\label{eq:realDiffusionConstant}
  \sigma=4(1-\rho),\,\,\,\mathcal{D}=\rho^{-1}-1,
\end{equation}
which agree with Monte-Carlo simulations
 (see~Fig.~\ref{fig:autocorr}).

\emph{Ballistic regime.--}
By introducing the~charge imbalance~$ \mu \neq 0 $, the~local
observables cease to be orthogonal, $ \langle \oo{2}\oo{1}\rangle_{p(\rho,\mu)}\neq 0$ .
However, we may still consider only a~subspace of observables, with at most
one occurence of~$\oo{2}$ and~single~$\oo{1}$. In this case the relevant  subspace is spanned by $\{ y_0,y_1,z_{2d}^0,z_{2d+1}^1;d\geq 0\} $, with the observables $ z_d^k $ defined as
\begin{equation}
  z_d^k=\sum_{x}([0 \underbrace{0 \ldots 0}_{k} 1 \underbrace{0\ldots 0}_d 2]_{2x}-[0 2\underbrace{0\ldots 0}_d 1 \underbrace{0\ldots 0}_k]_{2x}).
\end{equation}
Performing a similar calculation as above we obtain the following expression for the time autocorrelation function (Sect.~B of \cite{sup})
\begin{equation}\label{eq:autocorrZeroDelta}
  \frac{C(t)}{8\bar{\rho}}=\begin{cases}
    \mu^2+\rho(2-\rho);& t=0,\\
    \frac{2\mu^2}{\rho}+2\bar{\rho}^3(1-2\rho)^{2t-2}\left(\rho-\frac{\mu^2}{\rho}\right);& t\ge 1,
  \end{cases}
\end{equation}
with $ \bar{\rho}=1-\rho $, which immediately yields the~exact expression for the~Drude weight
\begin{equation}\label{eq:drudeWeight}
  D=16\left(\rho^{-1}-1\right)\mu^2,
\end{equation}
again agreeing excellently with numerics (Fig.~\ref{fig:autocorr}).

{\bf Inhomogeneous quench.--}
Let us consider dynamics of charges starting from the~product initial
state~$ p $ with uniform density~$ \rho $ and two
distinct chemical potentials $ \mu_L,\,\mu_R $ on the~left and the~right half
of the~chain, which can be expressed locally as
\begin{equation}
  p_{n/2 + x}=[0]'+\mu_{x} [1]',\ 
  \label{eq:inhstates}
\end{equation}
with $ \mu_x=\mu_L $ for $ x\leq 0 $ and $ \mu_R $ for $x>0$.
We shall now discuss the time evolution of the~centered charge profile
\begin{equation}
  f(x,t)=\langle U^t q_{n/2 + 2 x-1}\rangle_{ p} = \langle q_{n/2+2x-1}\rangle_{U^{-t} p}.
  \label{eq:cd}
\end{equation}
We are interested in the dynamics inside of the light-cone $ |x|\leq t $ with $ n\ge 8\, t+2 $, so that the boundary values of the profile are constant $ f(\mp t,t)=2\mu_{L,R} $.

Due to the~conservation of the~number of $\alpha_j=1$ in the adjoint time evolution $U^{-t} p$ expressed in the dual basis $[\un{\alpha}]'$, the initial state $p$
can be replaced by
\begin{equation}
\label{statey}
\tilde{p}=\sum_{x=-\frac{n}{2}+1}^{\frac{n}{2}}  \mu_{x} e_x,\ \ \text{where}\quad e_x = [\underbrace{0\ldots 0}_{\frac{n}{2}+x-1} 1\underbrace{0\ldots 0}_{\frac{n}{2}-x}]'.
\end{equation}

\begin{figure}
  \centering
  \includegraphics[width=\columnwidth]{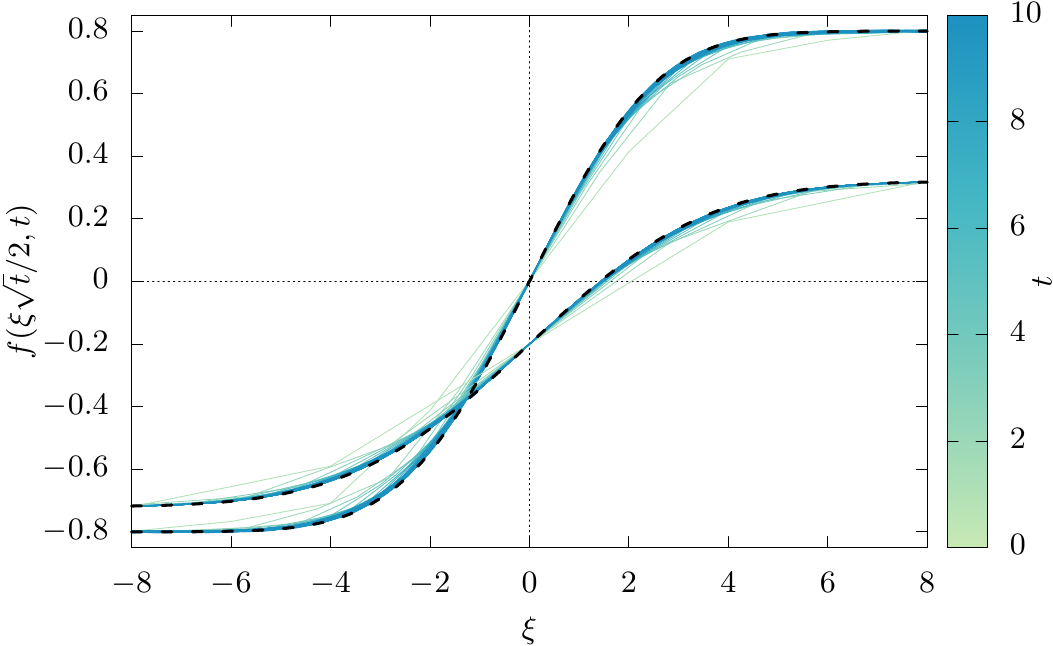}
  \caption{\label{fig:quench}
  Charge profiles~$f(\xi\sqrt{t}/2,t)$ after the~inhomogeneous
  quench. Curves with different colors correspond to different times~$t$.
  The profiles converge to the~estimated asymptotic profiles (dashed lines)
  given by Eq.~\eqref{f(v)}. The~parameters~$(\rho,\mu_1,\mu_2)$
  are~$(1/4,-0.36,0.16)$ and~$(1/3,-0.4,0.4)$.
}
\end{figure}
Here we use the dual space version of the argument used in the computation of correlation functions.
After half of the~time step
the~time propagated charge density (\ref{statey})
consists of the~terms including a~single occurrence of~$[1]'$ and
combinations of~$[1]'$ and~$[2]'$ on neighboring sites on the background of $[0\ldots 0]'$. Using the~argument
of freely propagating $[2]'$, namely $U_{1,2}[02]' = [20]'$, $U_{1,2}[20]'=[02]'$, 
we can conclude that the~terms containing $[2]'$ can be disregarded at all time steps due to the~orthogonality to
charge densities $q_x$. The~complete time propagation on the~relevant subspace, spanned by $\{ e_x \}$, is described by a cyclic block three-diagonal matrix
\begin{equation}
\label{Pqu}
 \mathcal{U}=\begin{bmatrix}
    \ddots&\ddots&\ddots&&\\
    &c&a&b&&\\
    &&c&a&b&\\
    &&&\ddots&\ddots&\ddots\\
  \end{bmatrix},
\end{equation}
 where $ a$, $b$, $c $ are $ 2\times 2 $ blocks. The projected initial state \eqref{statey} reads $\un{p} = \bigoplus_{x=1}^{n} \mu_{x-n/2}$.
Assuming that $n\ge 8\, t+2$, the charge profile (\ref{eq:cd}) inside of the light cone no longer depends on $n$.  Thus the limit  $n\to\infty$ can be applied and an 
infinite matrix $\mathcal{U}$ can be diagonalized using the block Fourier transform, yielding two bands of eigenvalues $\lambda_{1,2}(k)$, $k\in [-\pi,\pi]$ and Bloch eigenvectors
$\un{\omega}_{1,2}(k) = \bigoplus_{x\in\ZZ} \un{v}_{1,2}(k) e^{ik x}$ (see Sect.~C of \cite{sup} for details). Additionally note that $\ave{ q_{x} U^{-t} e_{y}} = 0$, if $|x-y| >2 t$, so the charge density profile can finally be expressed as a Fourier integral
\begin{equation}
\begin{split}
    \label{funk}
    f(x,t)=\sum_{y=x-t}^{x+t}&\int_{-\pi}^{\pi} \!\dd k\, e^{\ii k (x-y)}\mu_{2y}\cdot\\
    &\cdot\left( \lambda_2^t(k)\tilde{\alpha}_2(k)-\lambda_1^t(k) \tilde{\alpha}_1(k) \right),
\end{split}
\end{equation}
where $ \tilde{\alpha}_{1,2}(k)=\alpha_{1,2}(k) \left([1,1]\cdot\underline{v}_{1,2}(k)\right) $, and $\alpha_{1,2}(k)$ are
the coefficients expressing the vector $[1,1]$ in terms of Bloch vectors $v_{1,2}(k)$ (details in Sect. C of \cite{sup}).
We are interested in the~behavior of~$ f(x,t) $ on the~diffusive
lines~$2x=\xi\sqrt{t}$ in the~large time limit~$ \tilde{f}(\xi)=\lim_{t\to\infty}f(\xi\sqrt{t}/2,t) $.
In this limit, the~contribution from the~term proportional to
$\lambda_2(k)^t $  can be disregarded, since ${\rm sup}_k |\lambda_2(k)| <1 $. Additionally,
one should note that~$ \lambda_1(0)=1 $, so large $t$ asymptotics of (\ref{funk}) can be obtained by expanding 
$ \lambda_1(k) $ around $k=0$, $\lambda_1(k)^t\simeq e^{-\gamma k^2 t}$,
$\gamma=\frac{1-\rho}{4 \rho} $. Introducing a~new variable $ h=k \sqrt{t} $
the~integral \eqref{funk} can be calculated exactly in the scalling limit $x,t\to\infty$ with $x/\sqrt{t}$ fixed, yielding  
\begin{equation}
  \label{f(v)}
  \tilde{f}(\xi)=(\mu_R+\mu_L)+(\mu_R-\mu_L)\,\text{erf}\left(\frac{\xi}{4\sqrt{\gamma}}\right).
\end{equation}
The~agreement between the~exact and numerical result can be seen in~Fig~\ref{fig:quench}.
Since the~solution of diffusion equation
$\frac{\partial}{\partial t} f(x,t)=\mathcal{D} \frac{\partial^2}{\partial x^2}f(x,t) $ for a step initial data reads
$f(x,t)\sim\text{erf}(\tfrac{x}{\sqrt{4 \mathcal{D} t}})$,
the~diffusion constant is~$\mathcal{D}=4\gamma=\rho^{-1}-1$. 

Interestingly enough, our inhomogeneous quench does not excite any ballistic transport for the the specific class of initial states considered (\ref{eq:inhstates}) even for $\mu_L+\mu_R\neq 0$, 
unlike the linear response result which has a finite Drude weight for $\mu\neq 0$.
The reason is simple: 
inhomogeneous quench does not excite any imbalance of vacancy momentum, which is a conserved quantity.
However, in the perturbative 
linear-potential quench derivation of linear response coefficients (Sect.~A of \cite{sup}) one has a natural vacancy momentum bias which generates the Drude weight.
 
{\bf Discussion.--} We have studied transport properties of
a~simple reversible and deterministic cellular automaton. Despite its simplicity the~model exhibits a~large
variety of transport phenomena, including charge diffusion, and offers analytical handle on
the~calculation of transport coefficients as well as exactly solving
interesting initial value problems. Algebraic structure of the model allows to identify microscopic mechanisms behind its various transport regimes.
Specifically, the~ballistic
behavior of certain degrees of freedom can induce the~diffusive transport of
charge carriers, which are completely frozen in their absence.

Our results open many interesting questions. Firstly, it should be clarified whether the many body deterministic diffusion mechanism 
disclosed here applies to other integrable models, in particular to quantum lattice models such as $XXZ$. A promising idea in this direction is a formulation
of quantum transport in terms of classical-like soliton gas \cite{doyon2017soliton}. Secondly, our exactly solvable model could serve as a testbed for a precise verification of predictions of GHD. And thirdly, one may imagine various solvable generalizations of our model. For example, we can define a stochastic Markov chain model by introducing a tunnelling probability $\Gamma$ for a particle exchange, i.e. to modify the following matrix elements of the local propagator 
$U_{(\pm,\mp),(\mp,\pm)} = \Gamma$, $U_{(\pm,\mp),(\pm,\mp)}=1-\Gamma$.  Our analysis can be straightforwardly extended, resulting in the same value of the Drude weight as for the deterministic case $\Gamma=0$, while the Green-Kubo diffusion constant has a continuous dependence on $\Gamma$
\begin{equation}
  \mathcal{D}=(\rho^{-1}-1)\frac{1-\rho\Gamma}{1-\Gamma},
\end{equation}
and diverges at $\Gamma=1$. This implies that the physics of our model is robust against adding a small noise or dissipation. We conjecture that one could
similarly study quantization of the model and demonstrate continuity of the classical limit.

\smallskip

We acknowledge fruitful discussions with Bruno Bertini, Spyros Sotiriadis, Enej Ilievski, Lenart Zadnik and Marko \v Znidari\v c. The work has been supported by grants P1-0044 and N1--0025 of Slovenian Research Agency, and ERC grant OMNES.

\bibliography{dif_cl}

\clearpage
\begin{widetext}
\begin{center}
\textbf{ \large {\em Supplemental material:}\\
Diffusion in integrable deterministic interacting lattice systems}
\end{center}
\section{A: Linear response coefficients.}
In the first part we revisit the initial value problem, which yields the linear
response conductivity and Drude weight. 
The~protocol starts by applying a~kick (or a quench) with a linear-potential field $h$
\begin{equation}
K(h)= 1+h\sum_{x=1}^{n-1} (x-\tfrac{n}{2}) q_{x},\qquad\text{with}\quad h < \frac{1}{n},
\end{equation} 
to the
initial translationally invariant equilibrium state~$p$
\begin{equation}\label{eq:linearQuench}
\widetilde{p}=K(h)p,
\end{equation}
which induces a constant gradient of
charge $ q_x $,  
so that
$\ave{q_{n/2+x}}_{\tilde{p}} =2\mu+\chi h x$, where $ \chi$ is the static susceptibility.
After the~initial quench
the~system is left to evolve autonomously. The time dependent current, averaged over two neighboring temporal layers, now reads
\begin{equation}
\frac{1}{2} \langle j_{n/2}^{t+1/2}+j_{n/2}^{t-1/2}\rangle_{\tilde{p}}
  =\frac{h}{2}
  \sum_{x=1}^{n-1}(x-\tfrac{n}{2})
  \langle
  (j_{n/2}^{1/2}+j_{n/2}^{-1/2})\,q^{-t}_{x}
  \rangle_{p}.
\end{equation}
Taking into account the continuity equation~\eqref{eq:continuityeq} and discarding boundary
terms, we can alternatively write the above expression as
\begin{equation}
\frac{1}{2} \langle j_{n/2}^{t+1/2}+j_{n/2}^{t-1/2}\rangle_{\tilde{p}}=
\frac{h}{2}\sum_{t^\prime=0}^{t-1}\langle (j_{n/2}^{1/2}+j_{n/2}^{-1/2})J^{t^{\prime}+1/2-t}\rangle_p
+\frac{h}{2}\sum_{x=1}^{n-1}(x-\frac{n}{2})
\langle (j_{n/2}^{1/2}+j_{n/2}^{-1/2})q_x\rangle_p.
\end{equation}
The last term vanishes as a~consequence of a~more general property, stating that the current in the bulk is an odd function of time $t'$ 
\begin{equation}
  \sum_{x=1}^{n-1}(x-\frac{n}{2})\langle j_{n/2}^{{t'}} q_x\rangle_p
  =-\sum_{x=1}^{n-1}(x-\frac{n}{2})\langle j_{n/2}^{-{t'}} q_x\rangle_p,
\end{equation}
which follows from $U^{-{t'}}=S^{-1} U^{t'} S$.
Finally, by
noting $\langle{j_{n/2}^{1/2} J^{t'+1/2}}\rangle_p=\langle{j_{n/2} J^{t'}}\rangle_p$ and
$\langle j_{n/2} J^{t'}\rangle_p=\langle j_{n/2}^{-t'} J\rangle_p$, we obtain
\begin{equation}
 \frac{1}{2} \langle j_{n/2}^{t+1/2}+j_{n/2}^{t-1/2}\rangle_{\tilde{p}}=
  h\left(
  \frac{1}{2} \langle j_{n/2} J\rangle_p
  +\sum_{t^{\prime}=1}^{t-1}\langle j^{t^\prime}_{n/2} J\rangle_p
  +\frac{1}{2} \langle j^{t}_{n/2} J\rangle_p
  \right).
\end{equation}
Linear response conductivity $\sigma$ in the~main text~\eqref{prev} follows directly from
the~definition $\ave{j} = \sigma h$, after taking first the limit $n\to\infty$ and then the limit $t\to\infty$,
while the Drude weight can be interpreted as the~rate at which the~current in the~system increases with time.

Note that this quench~\eqref{eq:linearQuench} at $\mu\neq0$ also excites the gradient of density of vacancies,
which can
be understood by explicitly expressing $q_x\,p$,
\begin{equation}
  \begin{aligned}
    q_x\,p=\cdots  p_{x-1} &
    \left(
    -\frac{\mu}{1-2\rho} \left(\oo{0}_x'+\oo{2}_x' \right)
    +\rho \oo{1}_x'
    \right)
     p_{x+1}\cdots+\\+
    \cdots & p_{x} 
    \left(
    -\frac{\mu}{1-2\rho} \left(\oo{0}_{x+1}'+\oo{2}_{x+1}' \right)
    +\rho \oo{1}_{x+1}'
    \right)
     p_{x+2}\cdots.
  \end{aligned}
\end{equation}
Here~$p_x$ are local distributions $p_x=\oo{0}'+\mu \oo{1}'$, as defined in the~main text~\eqref{state}.
Due to momentum conservation of vacancies, this implies ballistic transport and finite Drude weight
unlike the kink shaped inhomogeneous quench.

\section{B: Autocorrelation functions.}
Here we present further details on the calculation of the~autocorrelation function.
We start by expressing the local two-site propagator in the basis  $\{\oo{0},\oo{1},\oo{2}\}$,
\begin{equation}\label{eq:localOperatorPropagator}
  \begin{gathered}
    U_{1,2}\oo{\alpha \alpha}=\oo{\alpha \alpha},\qquad
    U_{1,2}\oo{02}=\oo{20},\qquad U_{1,2}\oo{20}=\oo{02},\\
    U_{1,2}
    \begin{bmatrix}
      \oo{01}\\
      \oo{10}\\
      \oo{12}\\
      \oo{21}
    \end{bmatrix}
    =\begin{bmatrix}
      \rho& 1-\rho & \rho & -\rho\\
      1-\rho & \rho & -\rho & \rho\\
      1-\rho & -(1-\rho) & 1-\rho & \rho \\
      -(1-\rho) & 1-\rho & \rho & 1-\rho
    \end{bmatrix}
    \begin{bmatrix}
      \oo{01}\\
      \oo{10}\\
      \oo{12}\\
      \oo{21}
    \end{bmatrix},
  \end{gathered}
\end{equation}
where $\alpha\in\{0,1,2\}$.

As argued bellow, the dynamics
can be restricted to the subspace~$\mathcal{A}_J$
spanned by the basis
\begin{equation}
  \label{baza}
  \{y_0,y_1,z_{2d}^0,z_{2d+1}^1;\,d\ge 0\}.
\end{equation}
First of all, note that we can only consider the observables with a~single~$\oo{1}$,
forming the subspace~$\mathcal{A}^{(1)}$.
Let~$P$ be a~linear projector $P:\,\mathcal{A}^{(1)}\to \mathcal{A}_J$,
$P^2=P$ and $Q=1-P$. We then observe that for every $a\in\mathcal{A}^{(1)}$, $\ave{J Q a}_p=0$,
and that $Q(\mathcal{A}^{(1)})$ is invariant under $SU^{\text{e}}$.
Therefore the current autocorrelation function can be obtained
by restricting the dynamics to~$\mathcal{A}_J$, i.e.~considering
the reduced operator~$\mathcal{U}$,
\begin{equation}
  \mathcal{U}=P\,SU^{\text{e}}P=\begin{bmatrix}
    1-2\rho          & 2\rho &   0   \\
    0           &   0       &   0   \\
    -2(1-\rho)  &   1-2\rho      &   0   \\
    & & \rho & 1-\rho \\
    & & 1-\rho &\rho & 0 \\
    & & & & \rho & 1-\rho \\
    & & & & 1-\rho & \rho & 0 \\
    & & & & & & & \ddots \\
  \end{bmatrix}.
\end{equation}
To obtain the time propagation of the current density, the matrix~$\mathcal{U}$
should be applied to the~initial current density~$\underline{J}$ expressed in term of basis
elements $y_d$ and $z_d^k$
\begin{equation}
  \underline{J}=
  \begin{bmatrix}
    2,&
    1,&
    -1,&
    0,&
    0,&
    \cdots
  \end{bmatrix}^T.
\end{equation}
Vector~$\underline{o}$
\begin{equation}
  \label{lv}
  \underline{o}=
  \begin{bmatrix}
    2\rho(1-\rho),&
    \rho^2+\mu^2,&
    -\rho^2-\mu^2,&
    -2\mu^2,&
    -2\mu^2,&
    \cdots
  \end{bmatrix}
\end{equation}
incorporates the~overlaps between the~basis elements $y_d$ and $z_d^k$ and the
local current density w.r.t.~the state~$p$~\eqref{state}.
The~correlation function now reads
\begin{equation}
  C(t)=4(1-\rho)\,
  \underline{o}\ \mathcal{U}^{2t}\underline{J}.
\end{equation}
Splitting the overlap vector~$\underline{o}$ into parts corresponding to~$\mu=0$ and~$\mu\neq0$,
$\underline{o}=\underline{o}_D+\mu^2\underline{o}_B$, the~correlation function
decomposes into the $ \mu $ independent part~and the~contribution
proportional to~$\mu^2$.
The $ \mu $ independent part can be reduced to the $ 3\times3 $ block of the matrix $ \mathcal{U} $ corresponding to the basis elements $ y_0,y_1,z_0^0 $, as expressed in the main text \eqref{eq:diffusiveautocorrelation}.
The contribution  from $ \underline{o}_B $ can be obtained  
by applying $ \mathcal{U} $ to the~vector $\underline{o}_B $,
  \begin{equation}
    \underline{o}_B\mathcal{U}^{2t}=
    \begin{cases}
      \begin{bmatrix}
        0,&1,& -1, & -2, & -2, & \cdots
      \end{bmatrix}; &  t=0,\\      
      \begin{bmatrix}
        2(\rho^{-1}-1)\left(
        1-(1-\rho)(1-2\rho)^{2t-1}
        \right),&
        2-4(1-\rho)^2 (1-2\rho)^{2t-2},&
        -2,&-2,&-2,&\cdots
      \end{bmatrix}; &  t\ge 1,
    \end{cases}
  \end{equation}
which yields the expression for $C(t)$ from
the~main text~\eqref{eq:autocorrZeroDelta}.

\section{C: Inhomogeneous quench.}
The~complete expressions regarding inhomogeneous initial value problem are presented here.
The local propagator takes the following form in the dual basis~$\{\oo{0}^{\prime},\oo{1}^{\prime},\oo{2}^{\prime}\}$,
\begin{equation}\label{eq:localDualPropagator}
  \begin{gathered}
    U_{1,2}\oo{\alpha \alpha}^{\prime}=\oo{\alpha \alpha}^{\prime},\qquad
    U_{1,2}\oo{02}^{\prime}=\oo{20}^{\prime},\qquad U_{1,2}\oo{20}^{\prime}=\oo{02}^{\prime},\\
    U_{1,2}
    \begin{bmatrix}
      \oo{01}^{\prime}\\
      \oo{10}^{\prime}\\
      \oo{12}^{\prime}\\
      \oo{21}^{\prime}
    \end{bmatrix}
    =\begin{bmatrix}
      \rho& 1-\rho & 1-\rho & -(1-\rho)\\
      1-\rho & \rho & -(1-\rho) & 1-\rho\\
      \rho & -\rho & 1-\rho & \rho \\
      -\rho & \rho & \rho & 1-\rho
    \end{bmatrix}
    \begin{bmatrix}
      \oo{01}^{\prime}\\
      \oo{10}^{\prime}\\
      \oo{12}^{\prime}\\
      \oo{21}^{\prime}
    \end{bmatrix},
  \end{gathered}
\end{equation}
with~$\alpha\in\{0,1,2\}$. Using the local propagator, we can explicitly propagate
basis states~$e_x=[\ldots 0 1 0 \ldots]'$ to obtain 
the~$2\times 2$ blocks of the cyclic block tridiagonal matrix~$\mathcal{U}$~\eqref{Pqu},
\begin{equation}
a=\begin{bmatrix}
\rho^2 & \rho \left(1- \rho \right)  \\
\rho \left(1- \rho\right)  &  \rho^2 \\
\end{bmatrix}, \\
\qquad b=
\begin{bmatrix}
0 & 0 \\
\rho \left(1- \rho\right)  & \left(1-\rho \right){}^2 \\
\end{bmatrix},\;
\qquad  c=
\begin{bmatrix}
\left(1- \rho\right){}^2 & \rho \left(1- \rho \right) \\
0 & 0 \end{bmatrix}.
\end{equation}
The~solution of the eigenvalue equation $\mathcal{U} \un{\omega} = \lambda \un{\omega}$, within the Bloch ansatz $\un{\omega} = \bigoplus_{r}\exp(\ii k r)\underline{v}(k)$
reading explicitly
\begin{equation}
  a\,\underline{v}(k)+\exp(-\ii k)\,c\,\underline{v}(k)+\exp(\ii k)\,b\
 \underline{v}(k)-\lambda(k) \underline{v}(k)=0
\end{equation}
yields two bands and the corresponding eigenvectors
\begin{align}
  \lambda_{1,2}(k)&=\rho^2+(1-\rho)^2\cos{k} \pm \half (\mathrm{e}^{-\ii k}+1)(1-\rho)\sqrt{2 \mathrm{e}^{i k} \left((1- \rho )^2 \cos
(k)+ \rho  (\rho +2)-1\right),}\\
  \underline{v}_{1,2}(k)&=\left[\frac{\mathrm{e}^{-i k}}{2\rho} \left((1-\mathrm{e}^{i k}) (1- \rho)\pm \sqrt{2 \mathrm{e}^{i k} \left((1- \rho )^2 \cos
  (k)+ \rho  (\rho +2)-1\right)}\right),1\right].
\end{align}

We wish to express the vector
$\underline{e}_{2y-1}+\underline{e}_{2y}$, occurring in the expression for $ \underline{p} $ ($\un{e}_y$ is a unit infinite dimensional vector in with a single $1$ at place $y\in\ZZ$ and zeros elsewhere) in terms of the eigenvectors 
$\underline{\omega}_{1,2}(k)=\bigoplus_{r}\exp(\ii k r)\underline{v}_{1,2}(k)$ by solving the~following system of equations
\begin{equation}
\label{expan}
 \underline{e}_{2x-1}+\underline{e}_{2x}= \int_{-\pi}^{\pi} \dd k e^{-\ii k x}(\alpha_1(k) \underline{\omega}_1(k)+\alpha_2(k)\underline{\omega}_2(k)).
\end{equation}
Using the inverse discrete Fourier transform we obtain
\begin{equation}
  \label{soeq}
  \frac{1}{2\pi}\sum_{x=-\infty}^{\infty} e^{ik x}(\underline{e}_{2x-1}+\underline{e}_{2x})= (\alpha_1(k) \underline{\omega}_1(k)+\alpha_2(k)\underline{\omega}_2(k)).
\end{equation}
Expressing $ \underline{\omega}_{1,2}(k) $ in terms of local components $ \underline{v}_{1,2}(k) $ yields a set of equations
\begin{equation}
\frac{1}{2\pi}(\underline{e}_{2x-1}+\underline{e}_{2x})=\bigoplus_{r\in\ZZ} \delta_{r,y}(\alpha_1(k) \underline{v}_1(k)+\alpha_2(k)\underline{v}_2(k)),
\end{equation}
which admit the following solution:
\begin{equation}\label{konst}
\alpha_{1,2}(k)=\frac{1}{4\pi}\pm\frac{\rho-1 + \mathrm{e}^{i k} (\rho+1)}
{4\pi\sqrt{2 \mathrm{e}^{i k} \left((1- \rho )^2 \cos(k)+ \rho(\rho +2)-1\right)}}.
\end{equation}
The vectors $ \un{e}_{2x-1}+\un{e}_{2 x} $ expressed in eigenbasis \eqref{expan} are combined to form the initial state $ \un{p} $, which can then be propagated in time. As mentioned in the main text, for a given $ x $ and $ t $ in the expression for profile $ f(x,t) $, only the maximal light cone contributions from $ \un{p} $ can be taken into account. To compactly represent the function $ f(x,t) $ the time and space independent constants are joined together as  $ \tilde{\alpha}_{1,2}(k)=\alpha_{1,2}(k) \left([1,1]\cdot\underline{v}_{1,2}(k)\right) $.
Inserting the expressions \eqref{konst} into the main text equation~\eqref{funk},
and making an approximation to the eigenvalues discussed in the
main text, the~following expression is obtained
\begin{equation}
  f(\xi\sqrt{t}/2,t)=\frac{1}{\sqrt{t\,\gamma\,\pi}}\sum_{y=\xi\sqrt{t}/2-t}^{\xi\sqrt{t}/2+t}\mu_{2 y}\,
  \mathrm{e}^{-\frac{\left(y/\sqrt{t}- \xi/2\right)^2}{4 \gamma
  }}.
\end{equation}
Identifying the above sum with the Gaussian integral in the limit $t\to\infty$, we obtain the~final result~\eqref{f(v)}.
\end{widetext}
\end{document}